\documentclass[aps,prd,twocolumn,showkeys,superscriptaddress,groupedaddress,nofootinbib,floatfix]{revtex4}

\usepackage{graphicx}% Include figure files
\usepackage{dcolumn}% Align table columns on decimal point
\usepackage{bm}% bold math
\usepackage{aas_macros}
\usepackage{amsmath}	% Advanced maths commands
\usepackage{amssymb}	% Extra maths symbols
\usepackage{epsfig,amsmath,natbib}
\usepackage{color,subfigure}
\usepackage{xcolor}
\usepackage{mathrsfs,amssymb,amstext}
\usepackage{ulem}
\usepackage{url}
\usepackage{adjustbox}
\usepackage{threeparttable}
\usepackage{longtable}
\newcommand{\Mpc}{\mathrm{~km~s^{-1}~Mpc^{-1}}}

\begin{document}

%\preprint{APS/123-QED}

\title{The Hubble constant from the improved lens modeling of cluster-lensed supernova Refsdal with new spectroscopic redshifts and the Jackknife method} % 

\author{Yuting Liu\footnote{{yutingl@imu.edu.cn}}}
\affiliation{School of Physical Science and Technology, Inner Mongolia University, Hohhot 010021, China}
\affiliation{Institute of Astronomy and Physics, Inner Mongolia University, Hohhot 010021, China }
\author{Masamune Oguri\footnote{masamune.oguri@chiba-u.jp}} 
\affiliation{Center for Frontier Science, Chiba University,  Chiba 263-8522, Japan}
\affiliation{Department of Physics, Graduate School of Science, Chiba University, Chiba 263-8522, Japan}

\date{\today}% It is always \today, today,

\begin{abstract}
Time-delay cosmology offers an alternative approach to measuring the Hubble constant ($H_0$), which is distinct from the cosmic distance ladder and cosmic microwave background radiation methods. In this study, we present an improved strong lens mass modeling analysis of the cluster-lensed supernova Refsdal, incorporating the latest spectroscopic redshift data from the MUSE on the Very Large Telescope and the James Webb Space Telescope CANUCS/Technicolor. The robustness of our lens models is confirmed using the Jackknife method. From our analysis that considers four lens mass models with different assumptions on profiles of dark matter halos and external perturbations, we derive a constraint on the Hubble constant of $H_0=66.0\pm{4.3}\Mpc$ after combining best-fitted values of the four lens models.
\end{abstract}

\keywords{cosmological parameters: individual (Hubble constant) - gravitational lensing: strong - supernovae: individual (SN Refsdal) - galaxies: clusters: individual (MACS J1149.5+2223)}

\maketitle

%\tableofcontents

\section{Introduction}\label{sec:intro}

The Hubble constant ($H_0$) is a fundamental parameter in cosmology, representing the current expansion rate of the Universe. The accurate determination of 
$H_0$ is crucial for understanding the age, size, and ultimate fate of the Universe. However, a discrepancy between measurements from the cosmic microwave background (CMB) and local distance indicators, which is often referred to as ``Hubble tension'', poses a significant challenge in modern cosmology (e.g., \cite{2019NatAs...3..891V}).

Strong gravitational lensing  offers a powerful method to measure 
$H_0$ independently of the CMB and local distance indicators. When a massive galaxy or a cluster lies along the line of sight to a distant source, its gravitational field can bend and magnify the light from the source, creating multiple images with measurable time delays. Combining the time delay measurement with mass modeling the lensing object enables a direct measurement of $H_0$ \cite{1964MNRAS.128..307R}. This approach is sometimes called time-delay cosmography (see e.g., \cite{2024SSRv..220...48B} for a review).

There are many previous studies that demonstrate the effectiveness of this approach, especially at the galaxy-scale. For instance, H0  Lenses in COSMOGRAIL's Wellspring (H0LiCOW) project team analyzed six gravitationally lensed quasars to infer $H_0=73.3^{+1.8}_{-1.7}\Mpc$ assuming a flat $\Lambda$CDM cosmology \cite{2020MNRAS.498.1420W}. Recently, Time Delay COSMOlogy (TDCOSMO) project
team obtained $H_0=71.6^{+3.9}_{-3.3}\Mpc$ from the analysis of eight gravitationally lensed quasars with more conservative assumptions on the lens mass distribution \cite{2025arXiv250603023T}.
In addition, a new constraint on the Hubble constant of $H_0=73.22^{+5.95}_{-5.43}\Mpc$ was derived by combining the analysis of the lensed quasar RXJ1131$-$1231 with 42 binary black holes from GWTC-3 standard sirens \cite{2025ApJ...985L..44S}.

With the development of observational equipment and technology, strong gravitational lensing at the cluster scale has also been playing a significant role in the research of measuring the Hubble constant. 
For instance, the first system of a quasar lensed by a galaxy cluster SDSS J1004+4114 was investigated to infer $H_0=67.5^{+14.5}_{-8.9}\Mpc$ with the latest time-delay measurements and the combination of 16 different lens mass models in Ref. \cite{2023PhRvD.108h3532L}. 
Their analysis demonstrates that including more constraints on the lens potential is crucial for deriving precise constraints on the Hubble constant from cluster-lensed quasar systems (see also e.g., \cite{2022ApJ...926...86A,2023ApJ...959..134N}).
More recently,
the first measurement of the Hubble constant of $H_0=75.7^{+8.1}_{-5.5}\Mpc$ from the multiply-imaged type Ia supernova SN H0pe that was discovered by James Webb Space Telescope (JWST), which represents only the second example of the Hubble constant from any multiply-imaged supernovae, was presented in Ref. \cite{2025ApJ...979...13P}.

In this study, we focus on 
 the first gravitationally lensed supernova with resolved multiple
images, the supernova (SN) Refsdal in MACS J1149.5+2223, which has widely been used to constrain the Hubble constant.
Ref. \cite{2018ApJ...853L..31V} obtained a best value of
$H_0=64^{+9}_{-11}\Mpc$ by comparing the time-delay predictions from numerous models with the measured value derived by Ref. \cite{2015Sci...347.1123K} from very early data in the light curve of SN Refsdal. 
Later, Ref. \cite{2023Sci...380.1322K}
derived constraints on the Hubble constant from the detailed analysis of the SN Refsdal's reappearance, with $H_0=66.6^{+4.1}_{-3.3}\Mpc$ from two models that are most consistent with the observations. A similar constraint on the Hubble constant as well as other cosmological parameters was obtained in Ref. \cite{2024A&A...684L..23G}.
As indicated in Ref. \cite{2024hct..book..251T}, 
the data quality and modeling accuracy can significantly affect the precision of the Hubble constant.

While we cannot prove that no better model exists, the current best available models of MACS J1149.5+2223 are thought to be sufficiently precise and accurate based on various empirical tests. Recently, \citet{2025PhRvD.111l3506L} carefully examined the effect of mass-model assumptions on measuring the Hubble constant from the cluster-lensed SN Refsdal, employing 23 lens mass models. The weak dependence on the choice of lens mass models was concluded in this analysis \cite{2025PhRvD.111l3506L}, suggesting the robustness of the constraint on the Hubble constant from SN Refsdal thanks to many multiple images behind MACS J1149.5+2223. They obtain a value of $H_0=70.0^{+4.7}_{-4.9}\Mpc$ after combining results of the 23 lens mass models.

In observations, more data have recently been taken for this cluster. These new observations include VLT/MUSE observations of MACS J1149.5+2223 targeting a northern region of the cluster \cite{2024A&A...689A..42S} and the JWST observation by the CAnadian NIRISS Unbiased Cluster Survey (CANUCS), a JWST Cycle 1 GTO program targeting MACS J1149.5+2223, as well as JWST in Technicolor, a Cycle 2 follow-up GO program \cite{2025arXiv250621685S}.

Additionally, a new method to validate strong lens mass models was proposed by \citet{2025OJAp....8E.123N}, in which the Jackknife method is proposed as a new method to quantify the validity of cluster-scale mass models.

In this paper, we leverage the progress on observations and the model validation technique to derive the improved constraint on the Hubble constant from SN Refsdal. Our analysis use 114 multiple images from 37 systems, complementing and improving our previous work in Ref. \cite{2025PhRvD.111l3506L}.

This paper is organized as follows.
We describe observations and mass modeling in Sec. \ref{sec:obs and model}. In Sec. \ref{sec:method and results}, the Jackknife method and results are presented. We then describe constraints on the Hubble constant in Sec. \ref{sec:h values}.
We conclude in Sec. \ref{sec:summary}.

\section{Observations and Mass modeling}\label{sec:obs and model}

The gravitationally lensed SN Refsdal was discovered near the center of the galaxy cluster MACS J1149.5+2223 ($z=0.541$) and represents a significant milestone in the study of gravitational lensing and supernovae \cite{2015Sci...347.1123K}. It was first identified in 2014 using the Hubble Space Telescope (HST) as part of the Frontier Fields program, which aimed to explore the most distant galaxies in the Universe by utilizing the gravitational lensing effect of massive galaxy clusters.

The unique aspect of SN Refsdal is its multiple images, which were formed due to the gravitational lensing effect of the massive galaxy cluster MACS J1149.5+2223. This phenomenon, together with the transient nature of SN, allowed us to observe the same supernova event at different times and positions, providing a rare opportunity to study the time delay between the images \cite{2016ApJ...817...60T}. This time delay is crucial for measuring the Hubble constant.

The observations of SN Refsdal in MACS J1149.5+2223 were conducted using various instruments, including the Hubble Frontier
Field (HFF),  Cluster
Lensing And Supernova survey with Hubble (CLASH), as well as the Grism Lens-Amplified Survey from Space (GLASS)
\cite{2016ApJ...817...60T,2016ApJ...822...78G,2015ApJ...812..114T,2014ApJ...784..128S}.

To further enhance the dataset of this rare lensing cluster and refine the strong lensing model, an additional 5.5 hours observation with the deep Multi Unit Spectroscopic Explorer (MUSE) was conducted by the Very Large Telescope (VLT) of the European Southern Observatory (PI: A. Mercurio)  \cite{2024A&A...689A..42S}. These observations, carried out in 2022 and 2023, focused on a region in the northern part of the cluster and revealed several new multiple image systems with spectroscopic redshifts.

More recently, the CANUCS/Technicolor Data Release 1 provides extensive imaging and spectroscopic data for MACS J1149.5+2223, which includes JWST NIRCam imaging, NIRISS slitless spectroscopy, and NIRSpec prism multi-object spectroscopy \cite{2025arXiv250621685S}. This data release enhances the understanding of the stellar population parameters and lens models, offering a comprehensive view of the cluster, and allowing us to significantly expand the number of multiple image systems with spectroscopic redshifts. Moreover, they provide spectroscopic confirmation of numerous cluster members that were previously identified only photometrically.
In this paper, we utilize 114 multiple images from 37 systems, 28 of which have spectroscopic redshifts, by compiling both past and the latest observational results. Among these, 5 images are from SN Refsdal, 30 images are from the host galaxies of SN Refsdal, and the remaining images are from other background sources. Compared to our previous study \cite{2025PhRvD.111l3506L} for which we utilize 109 multiple images from 36 systems with 16 of them having spectroscopic redshifts, we include more multiple images with spectroscopic redshifts at higher redshifts, $z\sim 6.0$. The constraint on the northern part of the cluster is also significantly improved. Table~\ref{tab:mul} in the appendix summarizes all the multiple image information used in this paper. We also employ the latest observations of time delays and magnification ratios of multiple images of SN Refsdal in Ref. \cite{2023ApJ...948...93K}. In addition, we consider information about the positions, ellipticities, position angles, as well as luminosity
ratios relative to the brightest galaxy in MACS J1149.5+2223
of 173 cluster member galaxies.

Such rich and valuable observational constraints will be incorporated into our lens modeling. For the lens modeling of this galaxy cluster MACS J149.5+2223, we primarily consider three components: dark matter halos, member galaxies, and additional perturbations. For dark matter halos, we employ the Navarro-Frenk-White (NFW) profile or Pseudo-Jaffe Ellipsoid (PJE) profile to describe the mass distribution. For member galaxies, while the galaxy producing Einstein cross is described using the PJE profile, the remaining member galaxies are characterized using scaled PJE profiles. Furthermore, we considered multipole perturbations to enhance the modeling precision. Detailed model parameters and optimization procedures are found in Ref. \cite{2025PhRvD.111l3506L}.

In order to take account of the effect of the mass model assumption, our analysis incorporates a total of four lens models that are summarized in Table \ref{table:results}, which well covers the full range of the systematic error associated with the mass model assumption studied in Ref. \cite{2025PhRvD.111l3506L}.
For the model components in Table \ref{table:results}, {\tt anfw} represents the NFW  profile for the dark matter halo, while {\tt jaffe}  denotes the PJE  profile for the halo. The {\tt gals} represents the scaled PJE profile applied to the majority of member galaxies, and {\tt pert} indicates external shear and multipole perturbations of order $m$. The numerical value  following {\tt anfw} or {\tt jaffe}  indicates the number of halos incorporated in the model.
Specifically, in m1 $({\tt anfw4}+{\tt gals}+{\tt pert}+{\tt mpole}(\tt m=3))$, we employ four NFW profiles to describe the dark matter halos, utilize scaled PJE profile for the majority of member galaxies, and incorporate external shear and third-order multipole perturbations ($m=3$). Building upon m1, m2 $({\tt anfw4}+{\tt gals}+{\tt pert}+{\tt mpole}(\tt m=3,4))$ considers higher-order perturbations ($m=4$); m3 $({\tt jaffe4}+{\tt gals}+{\tt pert}+{\tt mpole}(\tt m=3))$ employs four PJE profiles to characterize the dark matter halos; and m4 $({\tt anfw2}+{\tt jaffe2}+{\tt gals}+{\tt pert}+{\tt mpole}(\tt m=3))$ implements  two NFW profiles and two PJE profiles simultaneously to describe the dark matter halos. Positional errors of each model listed in Table \ref{table:results} are determined so that the reduced $\chi^2$ becomes roughly equal to one.

All our lens modeling calculations are performed using the widely-used gravitational lensing analysis software GLAFIC 
\cite{2010PASJ...62.1017O,2021PASP..133g4504O} in the same manner as in Ref. \cite{2025PhRvD.111l3506L}.

\begin{figure*}
%\centering
\includegraphics[width=8.4cm]{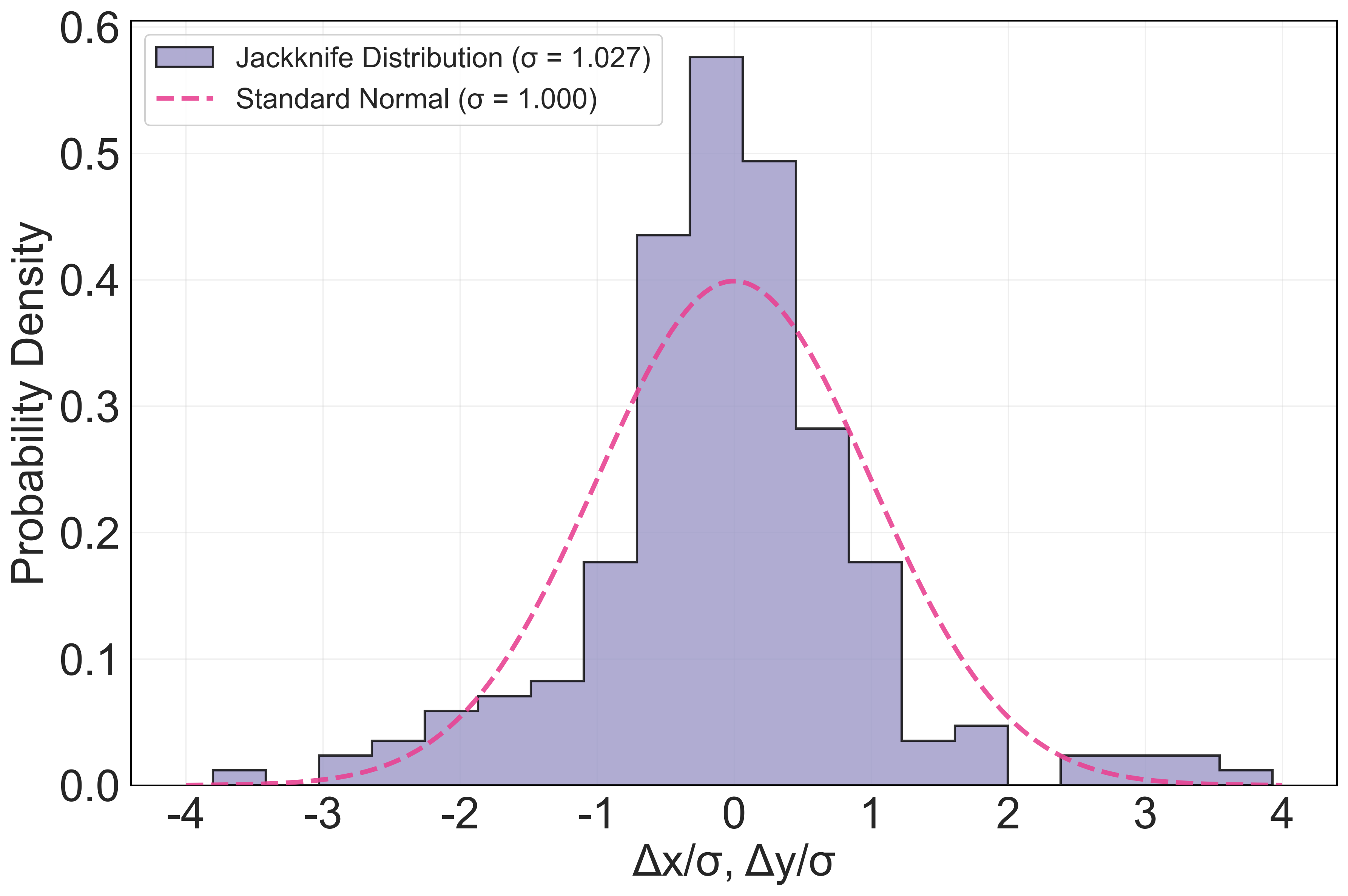}
\includegraphics[width=8.4cm]{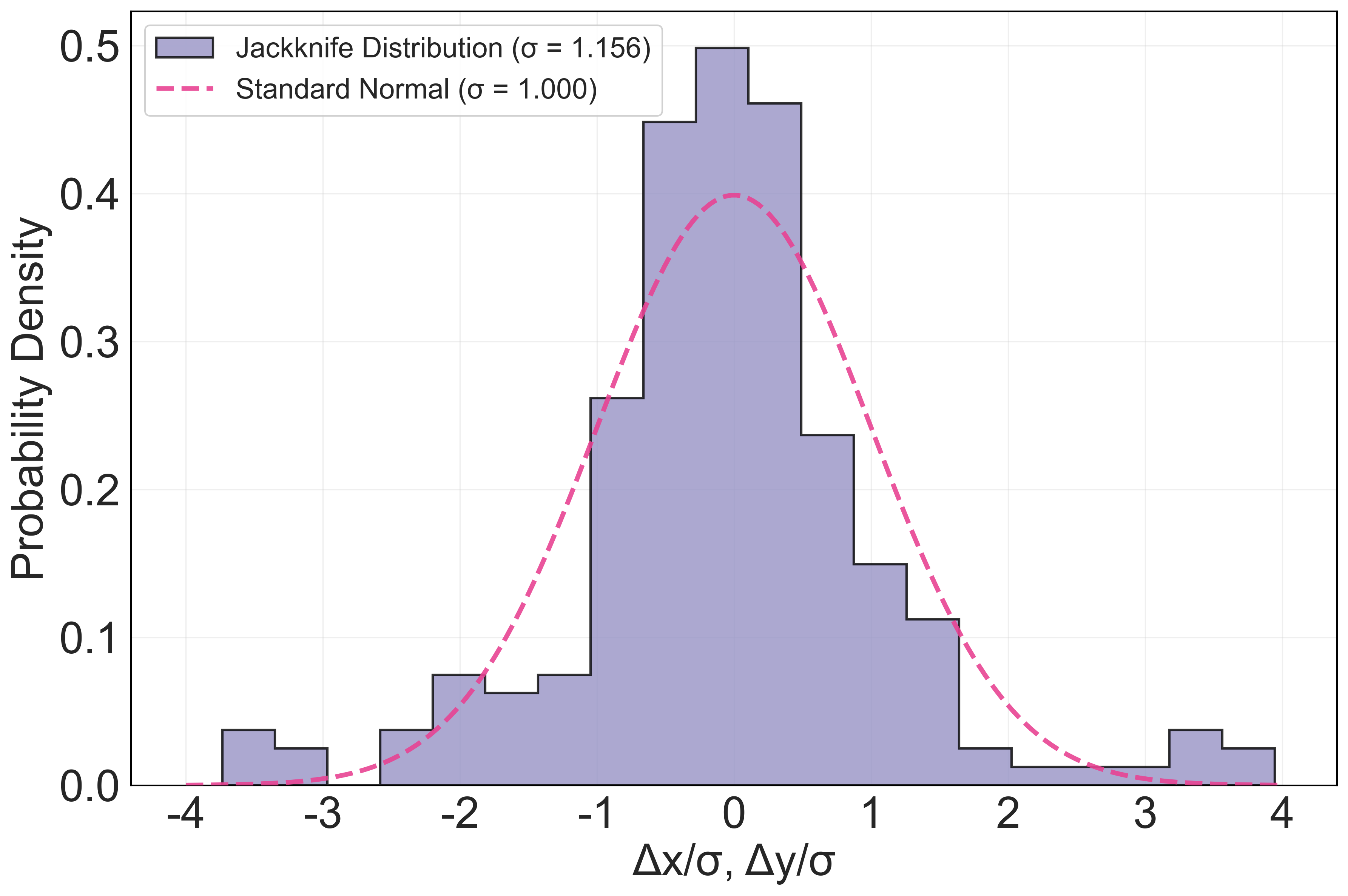}
\includegraphics[width=8.4cm]{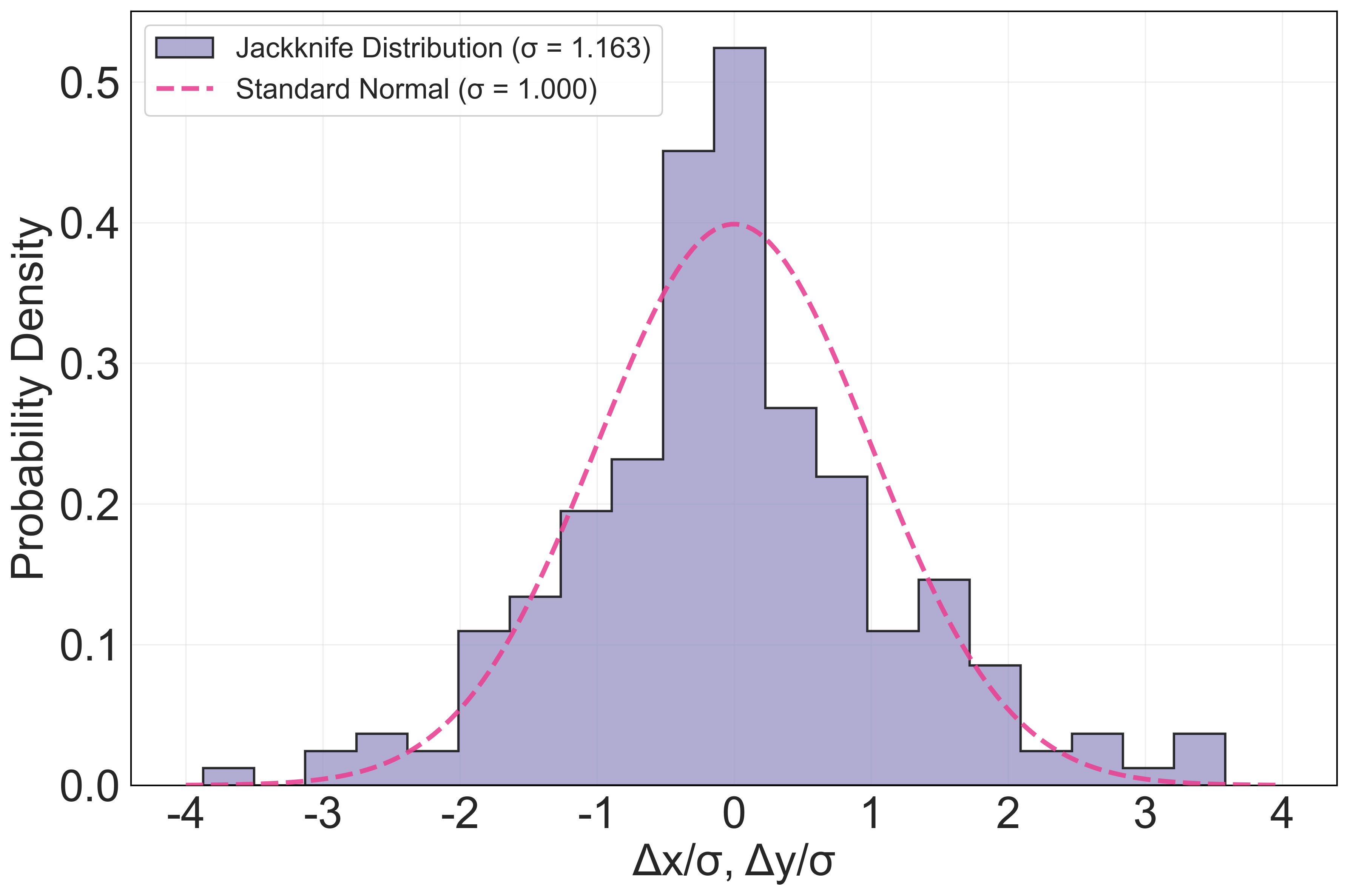}
\includegraphics[width=8.4cm]{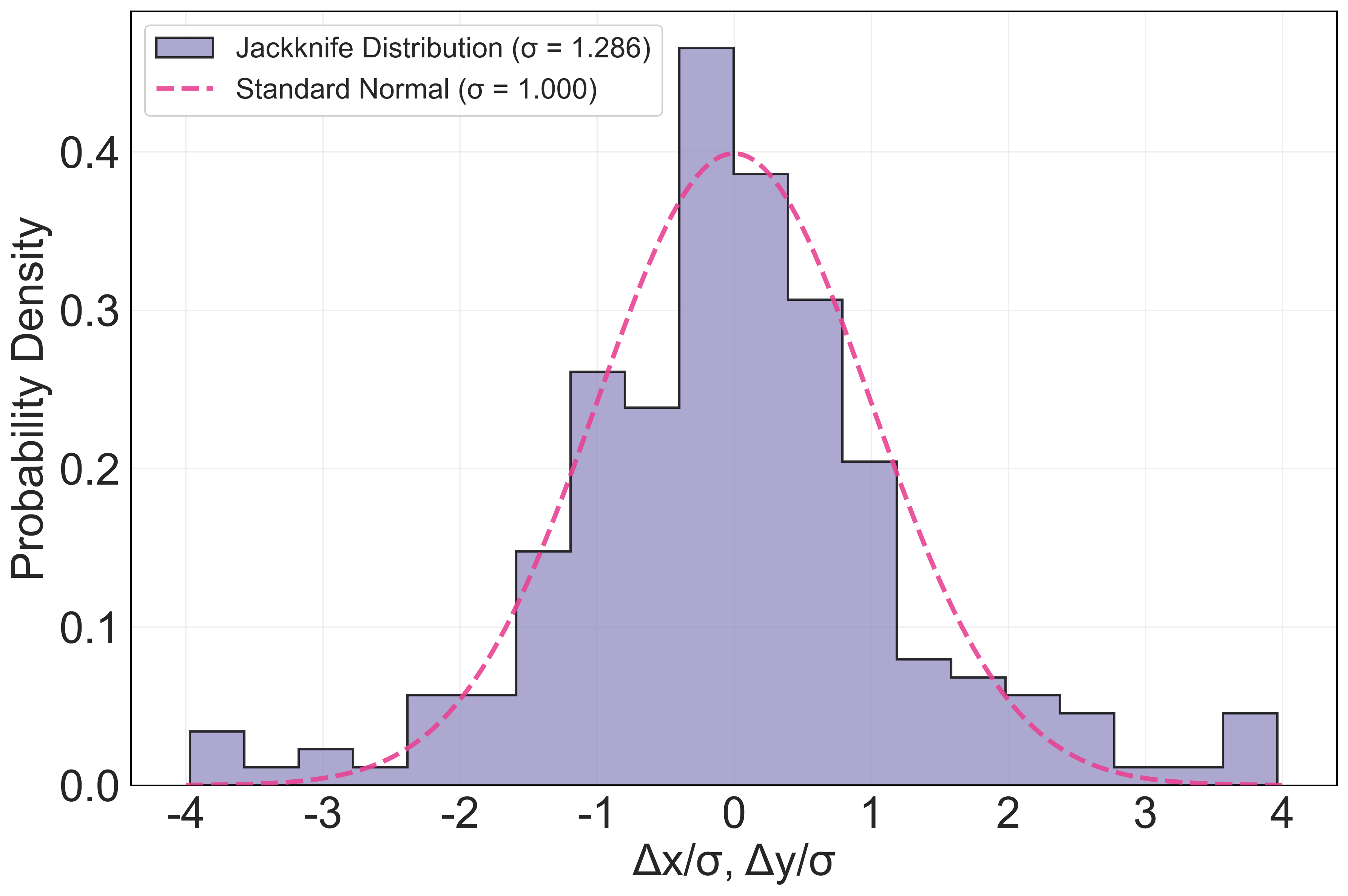}
\caption{The Jackknife distributions for all four different lens mass models with 4$\sigma$ outliers removed ({\it purple}), along with the standard normal distribution ({\it pink}). The upper panel from left to right are the model m1 and m2. The lower panel from left to right are the model m3 and m4.}
\label{Fig1}
\end{figure*}

\section{The Jackknife method and results}\label{sec:method and results}

The Jackknife method is a non-parametric method that systematically assesses the stability of an estimator by generating a series of sub-samples from the original dataset.
Given a dataset with $n$ observations, we construct $n$ Jackknife samples, each of which removes a single observation. For each sub-sample, the statistic of interest, $\hat{\theta}_{(-i)}$, is recalculated. The Jackknife estimate of the mean is then given by the average of these sub-sample estimates. The variance of the estimator is computed as

\begin{equation}
  \mathrm{Var}_{\mathrm{jack}}(\hat{\theta}) = \frac{n-1}{n} \sum_{i=1}^{n} \left( \hat{\theta}_{(-i)} - \bar{\theta}_{\mathrm{jack}} \right)^2,  
\end{equation}
where $\bar{\theta}_{\mathrm{jack}}$ is the mean of the Jackknife estimates. 

This approach provides a robust estimate of the statistical error without making assumptions about the underlying distribution of the data.
The Jackknife method is particularly useful for complex estimators or when the sample size is moderate, as it reduces bias and provides reliable error estimates. 

Since it was proposed in 1949, it has been widely applied in astrophysics and astronomy, especially for the large-scale structure analysis and power spectrum diagnostics \cite{2009MNRAS.396...19N,2011ApJ...739...85M,2014ApJ...784..128S,2020MNRAS.491.3290P}.
Recently, \citet{2025OJAp....8E.123N} extended its utility to the strong lens model validation, addressing limitations of the traditional 
$\chi^2$ validation when dark matter substructures introduce unquantifiable positional uncertainties.
The procedure is defined as follows:

\begin{enumerate}
    \item \textbf{Iterative source exclusion}: For a lens system with $R$  background sources, generate $R$ subsets by cyclically removing all multiple images of the $k$-th source ($k = 1, \dots, R$).
    
    \item \textbf{Model re-optimization}: For each subset, optimize mass model parameters $\boldsymbol{\theta}_{\rm lens}$ using the remaining $R-1$ sources:
    \begin{equation}
        \chi^2_{[k]} = \sum_{i \notin \mathcal{I}_k} \frac{\|\boldsymbol{\theta}^{\rm obs}_i - \boldsymbol{\theta}^{\rm model}_i(\boldsymbol{\theta}_{\rm lens})\|^2}{\sigma_{\rm eff}^2}
        \label{eq:chi2_jackknife}
    \end{equation}
    where $\mathcal{I}_k$ denotes images of the removed source, the $\sigma_{\rm eff}^2$ means the effective total positional uncertainty including the effect of substructures.
    
    \item \textbf{Prediction and residual analysis}: Using optimized parameters $\boldsymbol{\theta}_{\rm lens}^{[k]}$, compute predicted positions $\boldsymbol{\theta}^{\rm pred}_j$ ($j \in \mathcal{I}_k$) for the removed source. Calculate residuals as
    \begin{align}
        \Delta x_j &= x_j^{\rm obs} - x_j^{\rm pred}, \\
        \Delta y_j &= y_j^{\rm obs} - y_j^{\rm pred}.
        \label{eq:residuals}
    \end{align}

    \item \textbf{Statistical validation}: Aggregate normalized residuals ($\Delta x_j / \sigma_{\rm eff}, \Delta y_j / \sigma_{\rm eff}$) over all iterations. A robust model satisfies:
    \begin{itemize}
        \item \textbf{Gaussianity}: The residual distribution follows the Gaussian distribution.
        \item \textbf{Consistent scatter}: The standard deviation of the normalized residuals is $\sigma_{\rm Jackknife} \approx 1$.
    \end{itemize}
\end{enumerate}
This method has been successfully demonstrated on simulated lenses as well as the real cluster MACS0647 \cite{2025OJAp....8E.123N}.

In this paper, we follow this new method to validate our four lens models in MACS J1149.5+2223. The results are shown in Fig.~\ref{Fig1}.
We call
the distribution of $\Delta x/\sigma$, $\Delta y/\sigma$ a Jackknife distribution.
Overall, our Jackknife analysis reveals a robust mass models performance, with the distribution of positional offsets ($\Delta x/\sigma$, $\Delta y/\sigma$) closely following the standard normal curve $\mathcal{N}(0,1)$ and the standard
deviation of Jackknife distribution close to one. This analysis justifies our choice of positional errors shown in Table \ref{table:results}.

However, there are still some apparent differences in the Jackknife distributions among the four models.
For m1, we use the NFW profile to
describe four dark matter halos. The measured standard deviation $\sigma_{\rm Jackknife} = 1.027$ indicates a slight excess of 0. 03\% scatter beyond the ideal Gaussian distribution, which is most close to 1 in four models. Critically, the distribution does not show a significant skew or excess kurtosis, confirming the absence of systematic directional biases, while $>99.99\%$ of the residuals fall within $±4\sigma$, demonstrating effective rejection of catastrophic outliers.

For m2, we consider higher-order perturbations ($m=3,4$) compared to m1.
The Jackknife distribution ($\sigma_{\rm Jackknife} = 1.156$) represents a very slight change, which means that the addition of higher-order perturbations does not significantly affect our modeling results.

We also use the PJE profile to model four dark-matter halos, which corresponds to m3. The Jackknife distribution has undergone some changes compared with the results of m1, and the corresponding standard deviation is $\sigma_{\rm Jackknife} = 1.163$. This result is in line with expectations because using different profiles to describe the dark matter halo will yield slightly different results, which is also consistent with previous studies. 

Naturally, we also try using both the NFW profile and PJE profile simultaneously to describe the dark-matter halos, which corresponds to m4. It can be seen from Fig.~\ref{Fig1}  that there is no obvious change in the jackknife distribution. The standard deviation corresponding to m4 is the largest among the four models that deviates from 1, while this does not necessarily mean that m4 performs poorly in lens modeling,  because when we remove the 3$\sigma$ outliers instead of the 4$\sigma$ outliers the standard deviation corresponding to m4 is the closest to 1 among the four models. The results of the Jackknife analysis of the four models are summarized in Table \ref{table:results}.

\begin{figure*}
%\centering
\includegraphics[width=8.4cm]{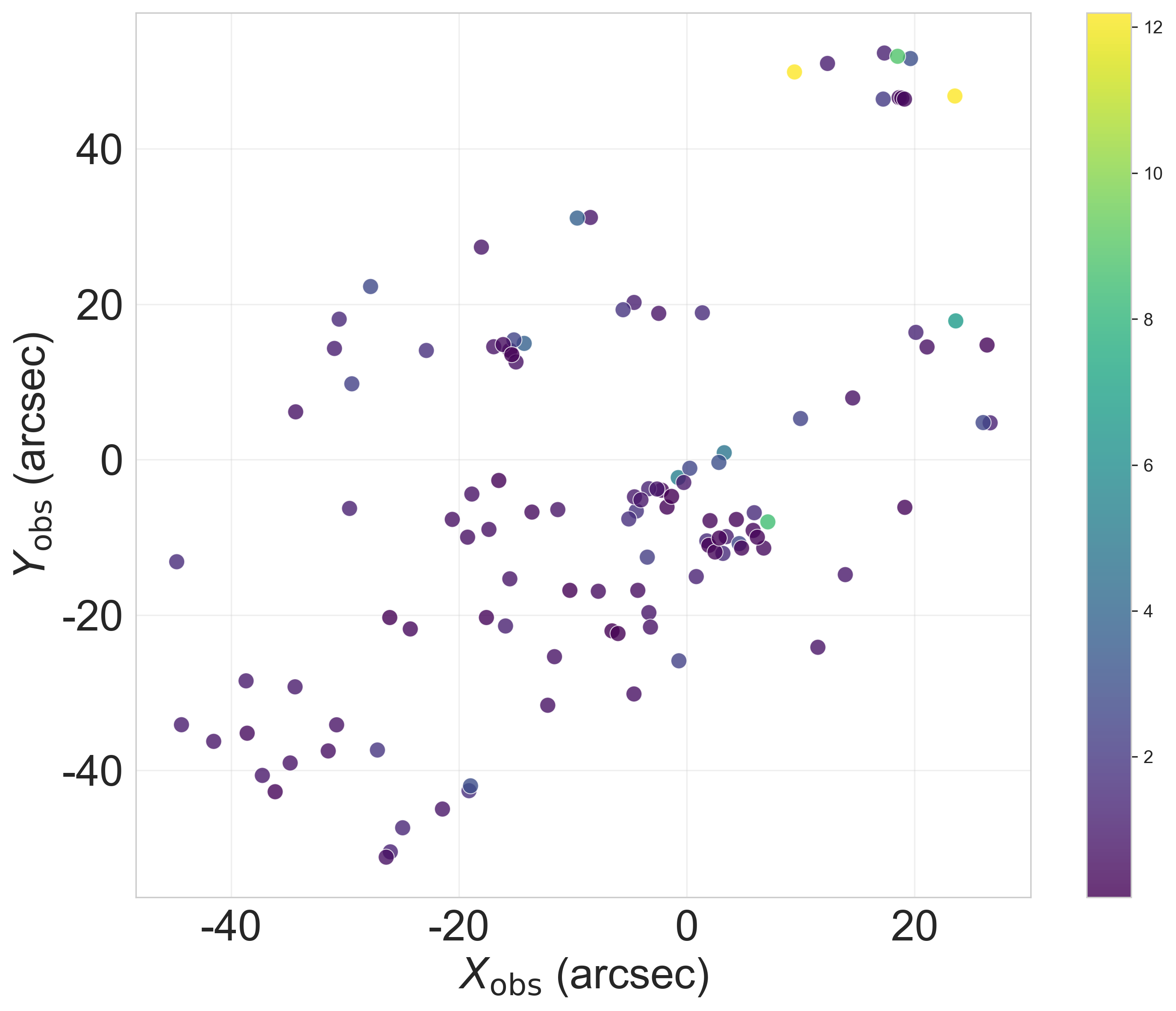}
\includegraphics[width=8.4cm]{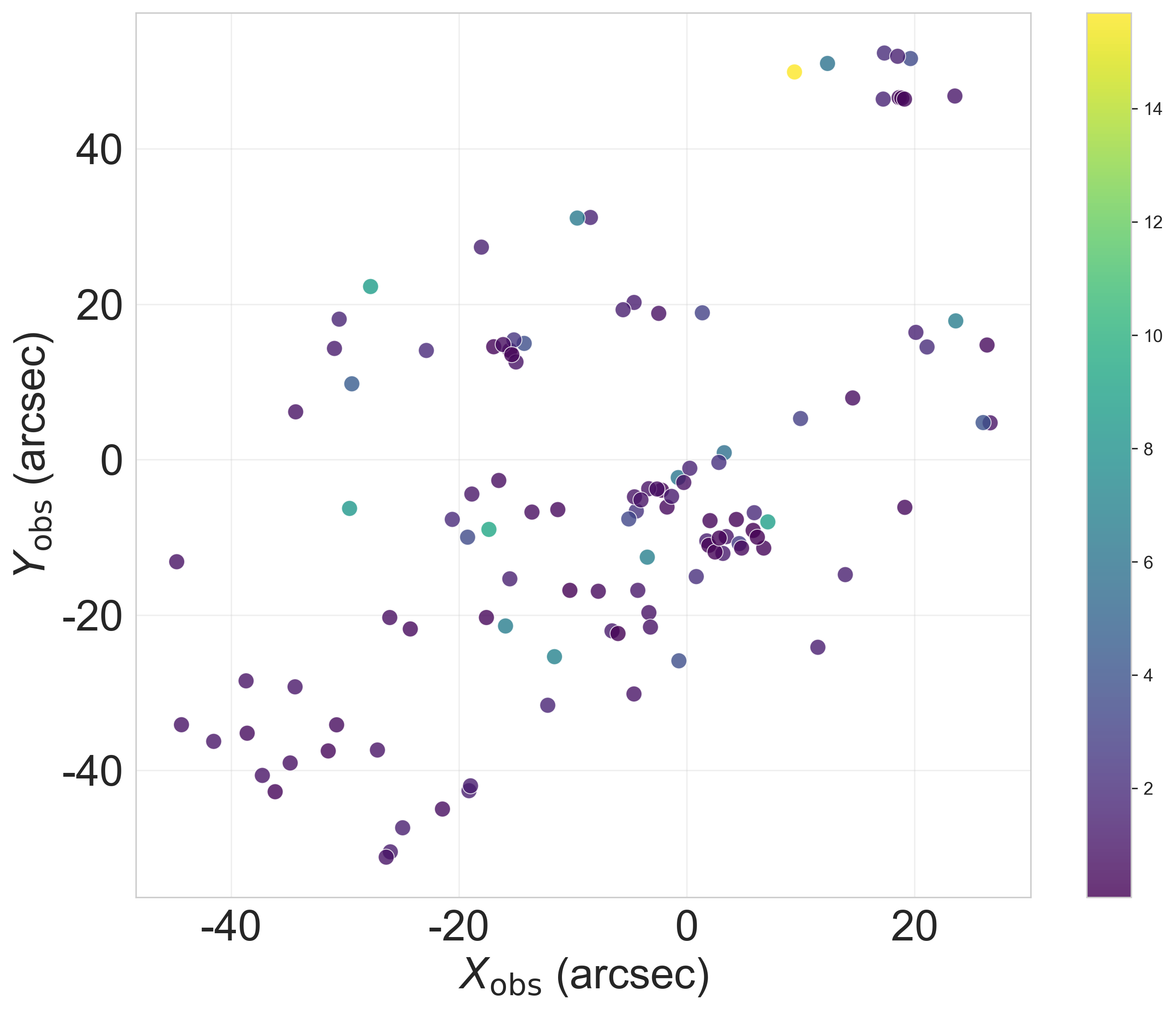}
\includegraphics[width=8.4cm]{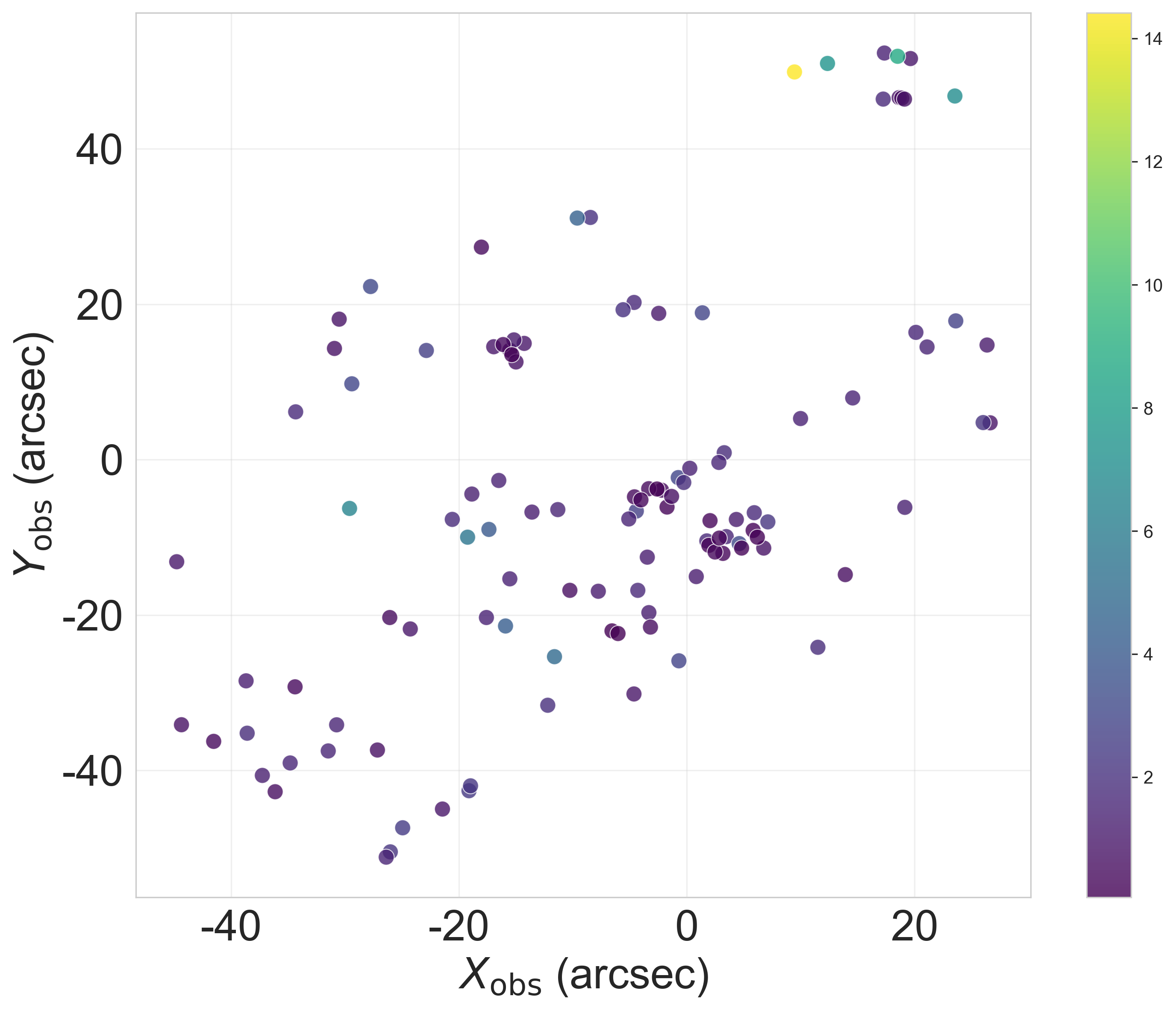}
\includegraphics[width=8.4cm]{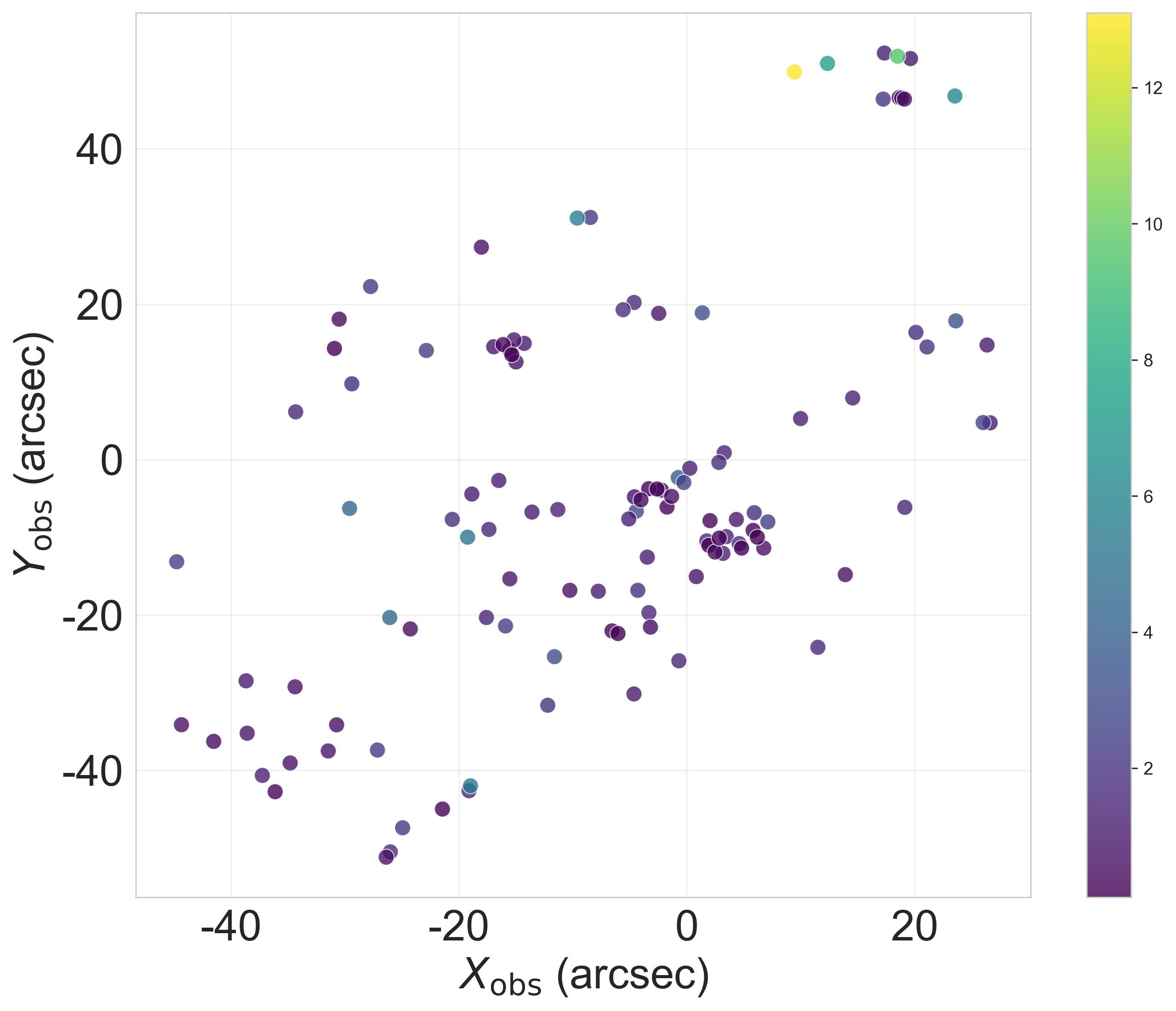}
\caption{The spatial distribution of $\Delta=\sqrt{(\Delta x/\sigma)^2+(\Delta y/\sigma)^2}$ for all 4 different lens mass models. The position ($X_{\rm obs}$, $Y_{\rm obs}$) refers to the image plane position with respect to the cluster center.
The upper panel from left to right are the model m1 and m2. The lower panel from left to right are the model m3 and m4.}
\label{Fig2}
\end{figure*}

To further validate our four lens models, we define $\Delta=\sqrt{(\Delta x/\sigma)^2+(\Delta y/\sigma)^2}$ for each multiple image.
In Fig. \ref{Fig2}, we present the spatial distribution of $\Delta$. We find that for the four lens models, the majority of $\Delta$ measurements performed well with their values close to 1. However, some minor trends emerge in some areas. In the region where $X_{\rm obs}<-20^{''}$ and $Y_{\rm obs}<-20^{''}$, and in the central region where $-10^{''}<X_{\rm obs}<10{''}$ and $-20^{''}<Y_{\rm obs}<0^{''}$, the $\Delta$ values fit very well. In the southwest region ($X_{\rm obs}>0^{''}$ and $Y_{\rm obs}>40^{''}$), few measurements perform poorly.

In our analysis,
the Jackknife method is helpful to evaluate our four lens models in MACS J1149.5+2223, effectively validating their reliability. The robust lens modeling is crucial for the application of accurate and precise cosmology, particularly in the measurement of the Hubble constant. With these four lens models, we will present new measurement results for the Hubble constant.

\section{Constraints on Hubble constant}\label{sec:h values}
\begin{table*}
  \caption{Summary of assumed positional errors of host knots images $\sigma_{\rm h}$ and galaxy images $\sigma_{\rm g}$, the standard deviation of the Jackknife distribution with 3$\sigma$ outliers removed $\sigma_{J_3}$ and  4$\sigma$ outliers removed $\sigma_{J_4}$, 
  the Hubble constant with 68.3$\%$ confidence interval, reduced-$\chi^2$ ($\chi^2$/dof) values, as well as label in Ref. \cite{2025PhRvD.111l3506L} for 4 different lens mass models.} 
  \label{table:results}
  %\vspace{2mm}
  %\begin{center}
  \begin{threeparttable}
  \setlength{\tabcolsep}{1.0mm}{
    \begin{tabular}{ccccccccc}
     \hline\hline
Label &  Model \tnote{*}   &  $\sigma_{\rm h}$[$''$]   & $\sigma_{\rm g}$[$''$] & $\sigma_{\rm J_3}$ & $\sigma_{\rm J_4}$ & $H_0$ [$\Mpc$]& $\chi^2$/dof & Label in Ref.\cite{2025PhRvD.111l3506L} \\\\
    \hline
m1 &{\tt anfw4}+{\tt gals}+{\tt pert}+{\tt mpole}(\tt m=3)   & 0.20 & 0.60 &0.910 & 1.027& $64.2 ^{+3.0}_{-3.2}$ &95.95/116 & M7\\
m2 &{\tt anfw4}+{\tt gals}+{\tt pert}+{\tt mpole}(\tt m=3,4)   & 0.20 & 0.50 & 0.930& 1.156 &$63.1^{+3.2}_{-3.2}$ &107.18/118 & M8 \\
m3  &{\tt jaffe4}+{\tt gals}+{\tt pert}+{\tt mpole}(\tt m=3) & 0.20 & 0.40 &1.053&1.163 & $68.3^{+3.2}_{-3.4}$ &111.70/121 & M17\\
m4 &{\tt anfw2}+{\tt jaffe2}+{\tt gals}+{\tt pert}+{\tt mpole}(\tt m=3) & 0.20 & 0.40 &1.009&1.286& $69.3^{+3.0}_{-3.2}$ &113.11/118 & M22\\ 
  \hline
  \hline
\end{tabular}}
%\begin{tablenotes}
%\footnotesize
%\item[*] {For model components, {\tt anfw} denotes the NFW profile for the halo, {\tt jaffe} denotes the PJE profile for the halo, {\tt gals} denotes the scaled PJE profile for most member galaxies, {\tt pert} refers to an external shear, and {\tt pert} refers to multipole perturbations of order m. The numerical value following {\tt anfw} or {\tt jaffe} indicates the number of halos. Please refer to the main text for a detailed explanation.}
%\end{tablenotes}
  %\end{center}
  
\end{threeparttable}
\end{table*}

\begin{figure}
%\centering
\includegraphics[width=1.0\linewidth]{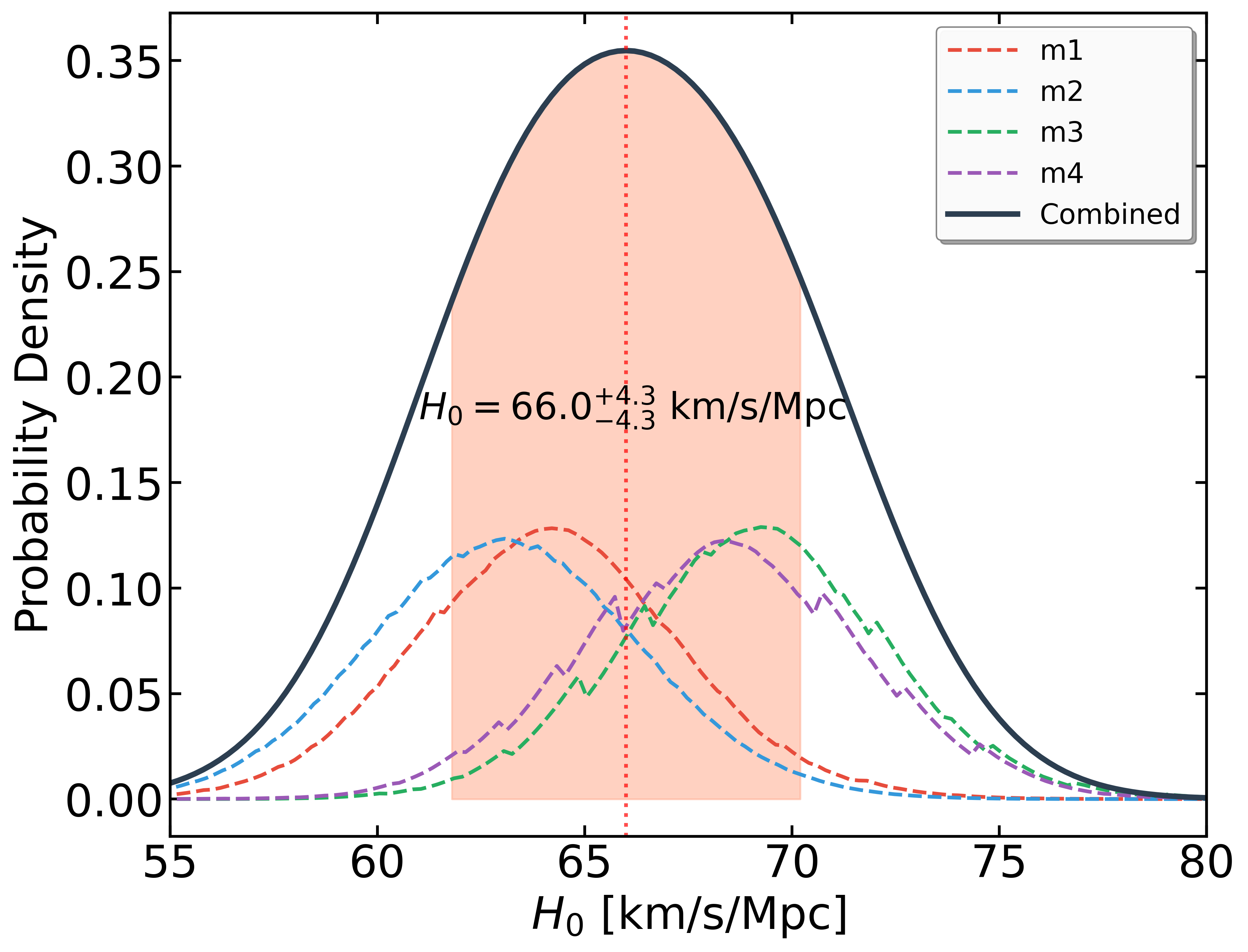}
\caption{The probability distribution functions (PDFs) for the Hubble constant, corresponding to each of the four distinct lens mass models, are illustrated with dashed lines. The solid line represents the PDF resulting from the equal-weighted combination of all the four models. The vertical orange dotted lines and the shaded area denote the median and the 68.3\% confidence interval of the combined PDF.
}
\label{Fig3}
\end{figure}

The time-delay cosmography method has emerged as a powerful tool in the precise determination of the Hubble constant ($H_0$), offering insights into the Universe's expansion history.  This technique leverages time delays between multiple images of gravitationally lensed objects and modeling the mass distribution of the lensing objects to infer the Hubble constant. 

Here we obtain new $H_0$
measurements derived from four improved lens models and follow the optimization method in our previous work in Ref. \cite{2025PhRvD.111l3506L}.
The results are summarized in Table \ref{table:results} and Fig. \ref{Fig3}.
We find that the differences in the Hubble constant measurements obtained from four different models are minimal, which is consistent with the conclusions reached in our previous work \cite{2025PhRvD.111l3506L}.

Moreover, the measured values of the Hubble constant are slightly smaller than those obtained from the corresponding models in previous work \cite{2025PhRvD.111l3506L}.
For instance, m1, which considers the matter distribution of four dark matter halos using the NFW profile, external shear, and multipole perturbations ($m=3$), yields a Hubble constant measurement of $64.2^{+3.0}_{-3.2}\Mpc$. This model corresponds to M7 in Ref. \cite{2025PhRvD.111l3506L}, in which the Hubble constant was measured as $66.9^{+3.6}_{-2.6}\Mpc$, slightly larger than the result of m1 in this paper.
Building on m1, we consider quadrupole perturbations ($m=3$, $4$) in m2, resulting in a Hubble constant measurement of $63.1^{+3.2}_{-3.2}\Mpc$, which is quite similar to the result of m1. 

When using the PJE profile instead of the NFW profile to describe the mass distribution of dark matter halos, as in m3, the Hubble constant measurement value changes to $68.3^{+3.2}_{-3.4}\Mpc$, which is larger than the result obtained from m1.
This indicates that different dark matter halo models can affect the measurement of the Hubble constant, which is also consistent with previous studies \cite{2023PhRvD.108h3532L,2025PhRvD.111l3506L}. 
For m4, where both the NFW profile and the PJE profile are used to describe the mass distribution of dark matter halos, the Hubble constant value obtained is $69.3^{+3.0}_{-3.2}\Mpc$. This represents a small change compared to m3, while a moderate change compared to m1.

In Fig. \ref{Fig3}, we present the posterior probability density distributions of the Hubble constant measurements obtained from the four improved models. By combining these distributions using an equal-weighted approach, we derive a combined Hubble constant measurement of $66.0^{+4.3}_{-4.3}\Mpc$.
We note that this measurement is very similar to the result of $66.6^{+4.4}_{-4.2}\Mpc$ reported in Ref. \cite{2023Sci...380.1322K} with  Oguri-a* and Grillo-g   models, demonstrating a strong consistency, even though in our new analysis we use many more multiple images with spectroscopic redshifts than those used in Ref. \cite{2023Sci...380.1322K}.

In general, the Hubble constant measurements obtained from our four improved models and the combined result are all less than $70\Mpc$, which aligns more closely with the result $67.4\pm0.6\Mpc$ obtained from the cosmic microwave background radiation \cite{2020A&A...641A...6P}.

\section{Conclusion}\label{sec:summary}

SN Refsdal in MACSJ1149+222.5 stands as the first multiply imaged supernova with measured time delays. This system merits detailed examination, as it offers valuable insights into the constraining power of cluster-scale strong lensing on the Hubble constant.

In this paper, we have updated our study on this rare cluster-lensed SN Refsdal system, incorporating the latest observations and the validation technique.
These include new VLT/MUSE observations targeting a northern region and new JWST observations by the CAnadian NIRISS Unbiased Cluster Survey (CANUCS) and JWST in Technicolor, providing NIRCam imaging in all wide and medium band filters over $\sim30$~arcmin$^2$ as well as NIRISS/NIRSpec spectroscopy. Our analysis incorporates these two recent observational datasets with the new list of multiple images summarized in Table \ref{tab:mul}. A total of 114 multi-images are included, from 37 systems, with the highest spectral redshift being 6.671. The number of multiple image systems with spectroscopic redshifts is nearly doubled compared with our previous analysis. In addition, we employ four lens mass models with varying assumptions regarding the density profiles of dark matter halos and external perturbation. All our mass models are validated with the Jackknife method that was recently proposed.

From the combination of all the four models with equal weighting, we have measured a value of $H_0=66.0\pm{4.3}\Mpc$.
Our finding is in excellent agreement with the result in Ref. \cite{2023Sci...380.1322K}. The value is slightly smaller than our previous result, $H_0=70.0^{+4.7}_{-4.9}\Mpc$, in Ref. \cite{2025PhRvD.111l3506L}, mainly because of the new spectroscopic redshifts particularly at high redshifts that better constrain the lens potential of the lensing cluster. Although the
Hubble tension remains unresolved by this work alone, our work shows that the
measurements are robust to modelling choices.
%Although the Hubble tension remains unresolved by this work alone, our measurements show a clear tendency toward values consistent with those derived from the cosmic microwave background radiation.

%A simple $\sqrt{N}$ scaling suggests that ten systems similar to MACS1149.5+2223 will suffice to reach $\sim2\%$ precision and thus contribute to solving the Hubble tension.

We note that our analysis does not take account of the mass-sheet degeneracy. Ref.~\cite{2020ApJ...898...87G} explicitly studied the effect of the external convergence for Refsdal to show that multiple images at different source redshifts constrain the value of external convergence effectively. Specifically, in Ref.~\cite{2020ApJ...898...87G}, assuming the mass sheet at the cluster redshift, the external convergence is constrained to [$-0.08$, $0.06$] at 1$\sigma$, which translates into $\sim 7\%$ error on the Hubble constant. Since the number of multiple images with spectroscopic redshift measurements is nearly doubled compared with those used in \cite{2020ApJ...898...87G}, we expect a much smaller error for our sample of multiple images. On the other hand, line-of-sight structures at other redshifts can also bias the Hubble constant value. We leave the detailed investigation of the line-of-sight effect to future work. 

Future observations will find many more cluster-scale strong lenses with time delay measurements \cite{2025OJAp....8E...8A} and will improve constraints on the Hubble constant further.

\begin{acknowledgments}
Y.L. was supported by the Natural
Science Foundation of Inner Mongolia of China (Grant
No.2025QN01042) and Research Start-up Fund of Inner Mongolia University (Grant No.10000-A25201021). M.O. was supported by JSPS
KAKENHI Grant Numbers JP25H00662, JP25H00672, JP22K21349.
\end{acknowledgments}

\appendix
\section{List of multiple images}\label{appendix:images}

We summarize the list of multiple images used in our analysis in Table~\ref{tab:mul}.

\begin{longtable}{ccccc}
\caption{\label{tab:mul}Multiple images used for the mass modeling. The multiple
    image set 2.1--2.5 is Supernova Refsdal, for which we also include flux
    ratios and time delays summarized in \cite{2025PhRvD.111l3506L} as
    observational constraints. The source redshifts $z_{\mathrm{s}}$
    with errors refer to photometric redshifts, for which we include
    the information as Gaussian priors. The superscript $^{a}$ denotes new
    spectroscopic redshifts that are not used in \cite{2025PhRvD.111l3506L},
    and multiple image sets 34--37 are also new. References before and after
    semicolons show those for multiple images and for spectroscopic
    redshifts, respectively. References; (1) -- \citet{2009ApJ...703L.132Z}; (2) -- \citet{2009ApJ...707L.163S}; (3) -- \citet{2015Sci...347.1123K}; (4) -- \citet{2023ApJ...948...93K}; (5) -- \citet{2014MNRAS.443..957R}; (6) -- \citet{2016MNRAS.457.2029J}; (7) -- \citet{2016ApJ...822...78G}; (8) -- \citet{2016ApJ...817...60T}; (9) -- \citet{2012Natur.489..406Z}; (10) -- \citet{2025arXiv250621685S}; (11) -- \citet{2016ApJ...819..114K}; (12) -- \citet{2014MNRAS.444..268R}; (13) -- \citet{2024A&A...689A..42S}}\\\hline\hline
\endfirsthead
\multicolumn{5}{c}{{\bfseries Table \ref{tab:mul} Continued}}\\\hline\hline
ID & RA &  Dec   & $z_{\mathrm{s}}$ & Ref.\\\hline
\endhead
\hline\hline
\endfoot
\endlastfoot
%\begin{ruledtabular}
  %\begin{tabular}{ccccc}
\hline\hline
ID & RA &  Dec   & $z_{\mathrm{s}}$ & Ref.\\
\hline
 1.1 & 177.397000 &  22.396000 &   $ 1.488$  & 1; 2 \\ 
 1.2 & 177.399417 &  22.397439 &             &  \\ 
 1.3 & 177.403417 &  22.402439 &             &  \\ 
\hline
 2.1 & 177.398222 &  22.395626 &   $ 1.488$  & 3, 4; 3 \\ 
 2.2 & 177.397710 &  22.395779 &             &  \\ 
 2.3 & 177.397368 &  22.395529 &             &  \\ 
 2.4 & 177.397797 &  22.395179 &             &  \\ 
 2.5 & 177.400081 &  22.396692 &             &  \\ 
\hline
 3.1 & 177.396612 &  22.396306 &   $ 1.488$  & 2; 2 \\ 
 3.2 & 177.398975 &  22.397889 &             &  \\ 
 3.3 & 177.397762 &  22.398778 &             &  \\ 
 3.4 & 177.398671 &  22.398222 &             &  \\ 
 3.5 & 177.403038 &  22.402686 &             &  \\ 
\hline
 4.1 & 177.398137 &  22.396350 &   $ 1.488$  & 2; 2 \\ 
 4.2 & 177.399271 &  22.396836 &             &  \\ 
 4.3 & 177.403842 &  22.402567 &             &  \\ 
\hline
 5.1 & 177.396721 &  22.395369 &   $ 1.488$  & 2; 2 \\ 
 5.2 & 177.399754 &  22.397492 &             &  \\ 
 5.3 & 177.400130 &  22.397201 &             &  \\ 
 5.4 & 177.403254 &  22.402022 &             &  \\ 
\hline
 6.1 & 177.396971 &  22.396633 &   $ 1.488$  & 2; 2 \\ 
 6.2 & 177.398829 &  22.397714 &             &  \\ 
 6.3 & 177.397904 &  22.398431 &             &  \\ 
 6.4 & 177.403308 &  22.402811 &             &  \\ 
\hline
 7.1 & 177.397443 &  22.396392 &   $ 1.488$  & 2; 2 \\ 
 7.2 & 177.399151 &  22.397217 &             &  \\ 
 7.3 & 177.403593 &  22.402644 &             &  \\ 
\hline
 8.1 & 177.396889 &  22.395759 &   $ 1.488$  & 5; 2 \\ 
 8.2 & 177.399535 &  22.397481 &             &  \\ 
 8.3 & 177.399959 &  22.397092 &             &  \\ 
 8.4 & 177.403364 &  22.402284 &             &  \\ 
\hline
 9.1 & 177.398168 &  22.395467 &   $ 1.488$  & 5; 2 \\ 
 9.2 & 177.398005 &  22.395229 &             &  \\ 
 9.3 & 177.397305 &  22.395369 &             &  \\ 
 9.4 & 177.397893 &  22.395726 &             &  \\ 
\hline
10.1 & 177.402417 &  22.389750 &   $ 1.894$  & 1; 2 \\ 
10.2 & 177.406042 &  22.392478 &             &  \\ 
10.3 & 177.406583 &  22.392886 &             &  \\ 
\hline
11.1 & 177.390750 &  22.399847 &   $ 3.129$  & 1; 6 \\ 
11.2 & 177.392708 &  22.403081 &             &  \\ 
11.3 & 177.401292 &  22.407189 &             &  \\ 
\hline
12.1 & 177.393000 &  22.396825 &   $ 2.949$  & 1; 7 \\ 
12.2 & 177.394375 &  22.400736 &             &  \\ 
12.3 & 177.404167 &  22.406128 &             &  \\ 
\hline
13.1 & 177.399750 &  22.393061 &   $ 2.800$  & 1; 8 \\ 
13.2 & 177.401083 &  22.393825 &             &  \\ 
13.3 & 177.407917 &  22.403553 &             &  \\ 
\hline
14.1 & 177.398500 &  22.394350 &   $\cdots$  & 1 \\ 
14.2 & 177.399792 &  22.395044 &             &  \\ 
14.3 & 177.407089 &  22.404719 &             &  \\ 
\hline
15.1 & 177.403708 &  22.397786 &   $ 1.240$  & 9; 8 \\ 
15.2 & 177.402833 &  22.396656 &             &  \\ 
15.3 & 177.400042 &  22.393858 &             &  \\ 
\hline
16.1 & 177.404034 &  22.392887 &   $ 3.214$  & 6; 6, 10 \\ 
16.2 & 177.400143 &  22.390154 &             &  \\ 
16.3 & 177.409064 &  22.400239 &             &  \\ 
\hline
17.1 & 177.393151 &  22.411472 &   $\cdots$  & 8 \\ 
17.2 & 177.393080 &  22.411456 &             &  \\ 
17.3 & 177.393009 &  22.411419 &             &  \\ 
\hline
18.1 & 177.399708 &  22.392544 &   $ 2.793^{a}$  & 1; 10 \\ 
18.2 & 177.401833 &  22.393858 &             &  \\ 
18.3 & 177.408042 &  22.402506 &             &  \\ 
\hline
19.1 & 177.398958 &  22.391339 &   $\cdots$  & 1 \\ 
19.2 & 177.403417 &  22.394269 &             &  \\ 
19.3 & 177.407583 &  22.401242 &             &  \\ 
\hline
20.1 & 177.392414 &  22.402559 &   $\cdots$  & 11 \\ 
20.2 & 177.401639 &  22.407167 &             &  \\ 
20.3 & 177.390935 &  22.399856 &             &  \\ 
\hline
21.1 & 177.391657 &  22.403491 &   $ 3.701$  & 12; 8 \\ 
21.2 & 177.390832 &  22.402628 &             &  \\ 
\hline
22.1 & 177.392851 &  22.412869 &   $ 3.447^{a}$  & 2; 13 \\ 
22.2 & 177.393539 &  22.413067 &             &  \\ 
22.3 & 177.395039 &  22.412694 &             &  \\ 
\hline
23.1 & 177.410343 &  22.388751 &   $ 3.189^{a}$  & 6; 10 \\ 
23.2 & 177.409209 &  22.387687 &             &  \\ 
23.3 & 177.406243 &  22.385369 &             &  \\ 
\hline
24.1 & 177.409080 &  22.390409 &   $ 2.640^{a}$  & 6; 10 \\ 
24.2 & 177.407984 &  22.389051 &             &  \\ 
24.3 & 177.404493 &  22.386694 &             &  \\ 
\hline
25.1 & 177.395293 &  22.391822 &   $ 2.950^{a}$  & 6; 13 \\ 
25.2 & 177.405618 &  22.402434 &             &  \\ 
25.3 & 177.402151 &  22.396744 &             &  \\ 
\hline
26.1 & 177.407643 &  22.396787 &   $\cdots$  & 8 \\ 
26.2 & 177.402239 &  22.391487 &             &  \\ 
26.3 & 177.403530 &  22.392584 &             &  \\ 
\hline
27.1 & 177.409943 &  22.387242 &   $ 4.732^{a}$  & 6; 10 \\ 
27.2 & 177.406568 &  22.384509 &             &  \\ 
27.3 & 177.411226 &  22.388459 &             &  \\ 
\hline
28.1 & 177.409605 &  22.386659 & $6.5\pm0.5$  & 6 \\ 
28.2 & 177.406676 &  22.384319 &             &  \\ 
28.3 & 177.412076 &  22.389054 &             &  \\ 
\hline
29.1 & 177.405193 &  22.386039 &   $ 3.408^{a}$  & 6; 10 \\ 
29.2 & 177.408205 &  22.388117 &             &  \\ 
29.3 & 177.410372 &  22.390622 &             &  \\ 
\hline
30.1 & 177.404418 &  22.397301 &   $\cdots$  & 8 \\ 
30.2 & 177.403968 &  22.396037 &             &  \\ 
\hline
31.1 & 177.404530 &  22.395759 &   $\cdots$  & 8 \\ 
31.2 & 177.404934 &  22.396394 &             &  \\ 
\hline
32.1 & 177.400722 &  22.392409 & $8.4\pm0.5$  & 11 \\ 
32.2 & 177.400564 &  22.392314 &             &  \\ 
\hline
33.1 & 177.412202 &  22.394881 &   $ 6.671^{a}$  & 11; 10 \\ 
33.2 & 177.404453 &  22.386869 &             &  \\ 
33.3 & 177.406908 &  22.388149 &             &  \\ 
\hline
34.1 & 177.400140 &  22.404150 &   $ 4.384^{a}$  & 13; 13 \\ 
34.2 & 177.398340 &  22.403780 &             &  \\ 
\hline
35.1 & 177.400280 &  22.396410 &   $ 4.497^{a}$  & 13; 13 \\ 
35.2 & 177.395750 &  22.400000 &             &  \\ 
35.3 & 177.394570 &  22.394420 &             &  \\ 
\hline
36.1 & 177.400430 &  22.403890 &   $ 5.806^{a}$  & 13; 13 \\ 
36.2 & 177.399490 &  22.403760 &             &  \\ 
\hline
37.1 & 177.395910 &  22.412390 &   $ 5.983^{a}$ & 13; 13 \\ 
37.2 & 177.393570 &  22.411420 &             &  \\ 
37.3 & 177.391680 &  22.411530 &             &  \\ 
37.4 & 177.393190 &  22.412950 &             &  \\ 
\hline
\end{longtable}

\bibliography{refer}% Produces the bibliography via BibTeX.

\end{document}